\documentclass[twocolumn,showpacs,preprintnumbers,amsmath,amssymb,superscriptaddress]{revtex4}

\usepackage{epsf}
\usepackage{graphicx}
\usepackage{sidecap}

\usepackage{color}

\def \beq {\begin{equation}}
\def \eeq {\end{equation}}
\begin{document}

\title{{Observation of topological nodal-line fermionic phase in GdSbTe}}


\author{M.~Mofazzel~Hosen}\affiliation {Department of Physics, University of Central Florida, Orlando, Florida 32816, USA}


\author{Gyanendra Dhakal} 
\affiliation {Department of Physics, University of Central Florida, Orlando, Florida 32816, USA}

\author{Klauss~Dimitri}\affiliation {Department of Physics, University of Central Florida, Orlando, Florida 32816, USA}

\author{Pablo Maldonado}
\affiliation {Department of Physics and Astronomy, Uppsala University, P.\,O.\ Box 516, S-75120 Uppsala, Sweden}

\author{Alex Aperis}
\affiliation {Department of Physics and Astronomy, Uppsala University, P.\,O.\ Box 516, S-75120 Uppsala, Sweden}

\author{Firoza Kabir}
\affiliation {Department of Physics, University of Central Florida, Orlando, Florida 32816, USA}




\author{Peter M.\ Oppeneer}
\affiliation {Department of Physics and Astronomy, Uppsala University, P.\,O.\ Box 516, S-75120 Uppsala, Sweden}



\author{Dariusz Kaczorowski}
\affiliation {Institute of Low Temperature and Structure Research, Polish Academy of Sciences,
	50-950 Wroclaw, Poland}

\author{Tomasz~Durakiewicz}
\affiliation {Condensed Matter and Magnet Science Group, Los Alamos National Laboratory, Los Alamos, NM 87545, USA} 
\affiliation {Institute of Physics, Maria Curie - Sklodowska University, 20-031 Lublin, Poland}

\author{Madhab~Neupane*}
\affiliation {Department of Physics, University of Central Florida, Orlando, Florida 32816, USA}

\date{18 June, 2013}
\begin{abstract}

{Topological Dirac semimetals with accidental band touching between conduction and valence bands protected by time reversal and inversion symmetry are at the frontier of modern condensed matter research. Theoretically one can get Weyl and/or nodal-line semimetals by breaking either one of them. Most of the discovered topological semimetals are nonmagnetic i.e. respect time reversal symmetry. Here we report the experimental observation of a topological nodal-line semimetallic state in GdSbTe using angle-resolved photoemission spectroscopy. Our systematic study reveals the detailed electronic structure of the paramagnetic state of GdSbTe. We observe the presence of  multiple Fermi surface pockets including a diamond-shape, an elliptical shape, and small circular pockets around the zone center and high symmetry M and X points of the Brillouin zone (BZ), respectively. Furthermore, we observe the presence of a Dirac-like state at the X point of the BZ. Interestingly, our experimental data show a robust Dirac like state both below and above the magnetic transition temperature (T$_N$ $\sim$ 13 K). Having relatively higher transition temperature (T$_N$ $\sim$ 13 K), GdSbTe provides an archetype platform to study the interaction between magnetism and topological states of matter.}

\end{abstract}
\date{\today}
\maketitle


The discovery of 3D topological insulator turned out to be enthralling achievement of the last decade \cite{Hasan, SCZhang}. A three-dimensional (3D) Z2 topological insulator (TI) is described as a crystalline solid that acts as a traditional insulator in the bulk but on its surface it has Dirac electron states that may be conducting, gapless, or even spin polarized \cite{Hasan, SCZhang, Hasan_review_2, Xia, Neupane_4}. The discovery of such a material dates to 2007 in a bismuth based compound initiating a domino effect in which the discovery of one exotic state led to another broadening our knowledge of quantum matter and bringing forth an army of researchers focused on this novel field  \cite{Hasan, SCZhang, Hasan_review_2, Xia, Neupane_4}. Now we have been able to predict a multitude of new materials with exotic states from topological crystalline insulators to topological Kondo insulators, 2D quantum spin Hall insulators and even topological superconductors using tools, such as symmetries. Consideration of electronic states protected by time-reversal, crystalline, and particle hole symmetries has led to the prediction of many novel 3D materials, which can support Weyl, Dirac, and Majorana fermions, and to new types of insulators such as topological crystalline insulators and topological Kondo insulators, topological superconductors, as well as 2D quantum spin Hall insulators with large band gaps capable of surviving room-temperature thermal excitations \cite{Hasan_review_2, TCI, Dai_LiFeAs, Neupane, MH2, MH4, Neupane_1}. Recent attention has focused on exploring non-trivial topological behavior in metals and semimetals without bulk band gaps, which can support symmetry protected gapless points in the Brillouin zone \cite{Dai, Neupane_2, Nagaosa, NdSb1, Young_Kane, TaAs_theory,TaAs_theory_1,Suyang_Science,Hong_Ding,Ilya_PRL, Hasan_2, Moore, Zhang}.

Topological materials offer transformational opportunities by providing platforms for entirely new classes of fundamental science studies of quantum matter as well as the development of new paradigms for the design of devices of the future for wide-ranging applications in next generation electronics, communications and energy technologies \cite{Hasan, SCZhang}. It is envisaged that novel quantum materials will render another breakthrough in technology development. Essentially currently available topological materials are mostly non-magnetic, while the magnetic topological materials are limited \cite{Cava1, Cava2, Cava3, CST}. There is an especially urgent need therefore to find new magnetic topological materials so that the unique potential of these materials for fundamental science studies and applications can be explored. This new realm may provide us new exotic states which could be helpful to revolutionize the technology to new height. This motivates us to study a magnetic material focusing on the topology and magnetism.  


\begin{figure*}
	\centering
	\includegraphics[width=0.9\linewidth]{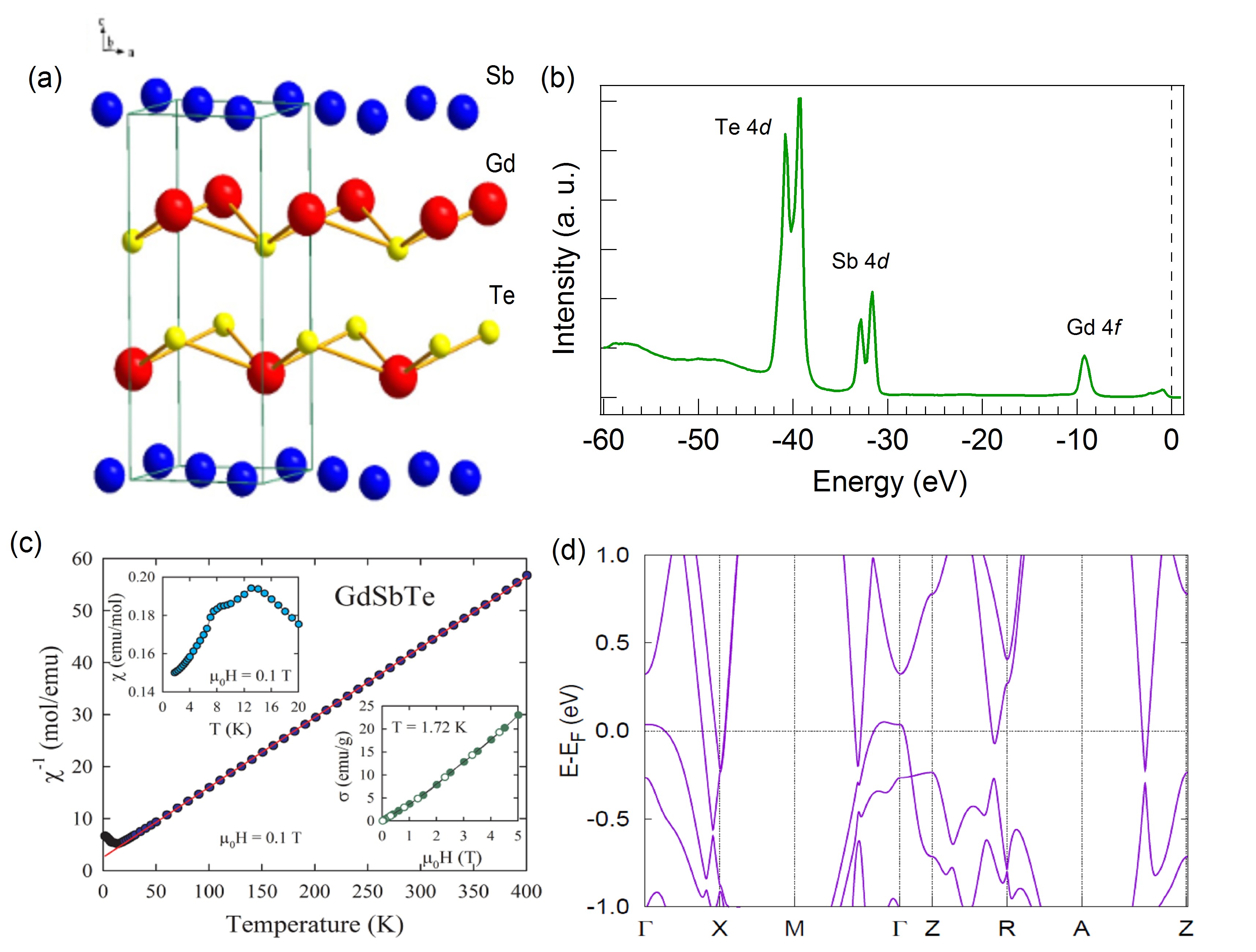}
	\caption{{Crystal structure and sample characterization of GdSbTe.} {(a)} Tetragonal type crystal structure of GdSbTe. Layers of Sb atoms form a square net around Gd atoms, which are separated by two Te layers. (b) Spectroscopically measured core-level of GdSbTe. Here, we clearly observed the sharp peaks of Te 4\textit{d} ($\sim$ 40 eV), Sb 4\textit{d} ($\sim$ 33 eV) and Gd 4\textit{f} ($\sim$ 8.5 eV). (c) Temperature dependence of the reciprocal magnetic susceptibility of single-crystalline GdSbTe measured in a magnetic field of 0.1 T applied along the crystallographic c-axis. Solid line represents the fit of Curie-Weiss law to the experimental data. Upper inset: low-temperature magnetic susceptibility data. Lower inset: magnetic field variation of the magnetization in GdSbTe measured at 1.72 K with increasing (full circles) and decreasing (open circles) magnetic field strength. (d) \textit{Ab-initio} calculated bulk band structure of GdSbTe along the various high-symmetry directions.}
\end{figure*}

In this report, we present a systematic studies of a magnetic material GdSbTe in order to reveal the topology and magnetism in this material using angle-resolved photoemission spectroscopy (ARPES), transport measurements and first-principles calculations. GdSbTe is isostructural with the nodal line semimetal ZrSiS \cite{Schoop,Neupane_5}. ZrSiS and related compounds have been shown to exhibit a diamond shaped Dirac line node as well as four-fold degenerate nodes at the high-symmetric points of the Brillouin zone (BZ) \cite{node_0, node_1, node_2, node_3, node_4, Schoop, Neupane_5, new_ZST, HfSiS, ZrGeM, ZrSiX, ZrSiTe, MSiS}. GdSbTe has a relatively higher magnetic transition temperature T$_N$, and can be accessed by ARPES and other techniques.
Our systematic electronic structure studies reveal the presence of multiple Fermi surface pockets such as a diamond, a circular and a elliptical shaped pockets around the $\Gamma$, X, and M point, respectively, of the Brillouin zone (BZ). Our study shows that GdSbTe system is not only structurally but also electronically similar to ZrSiS and related nonsymmorphic Dirac materials. Importantly, we observed a Dirac like state at the X point in both the paramagnetic and antiferromagnetic states of this material. Our findings provide a platform for discovering new exotic quantum phases evolved due to the interplay of magnetism with the topological phases.

\begin{figure*}
	\centering
	\includegraphics[width=18cm]{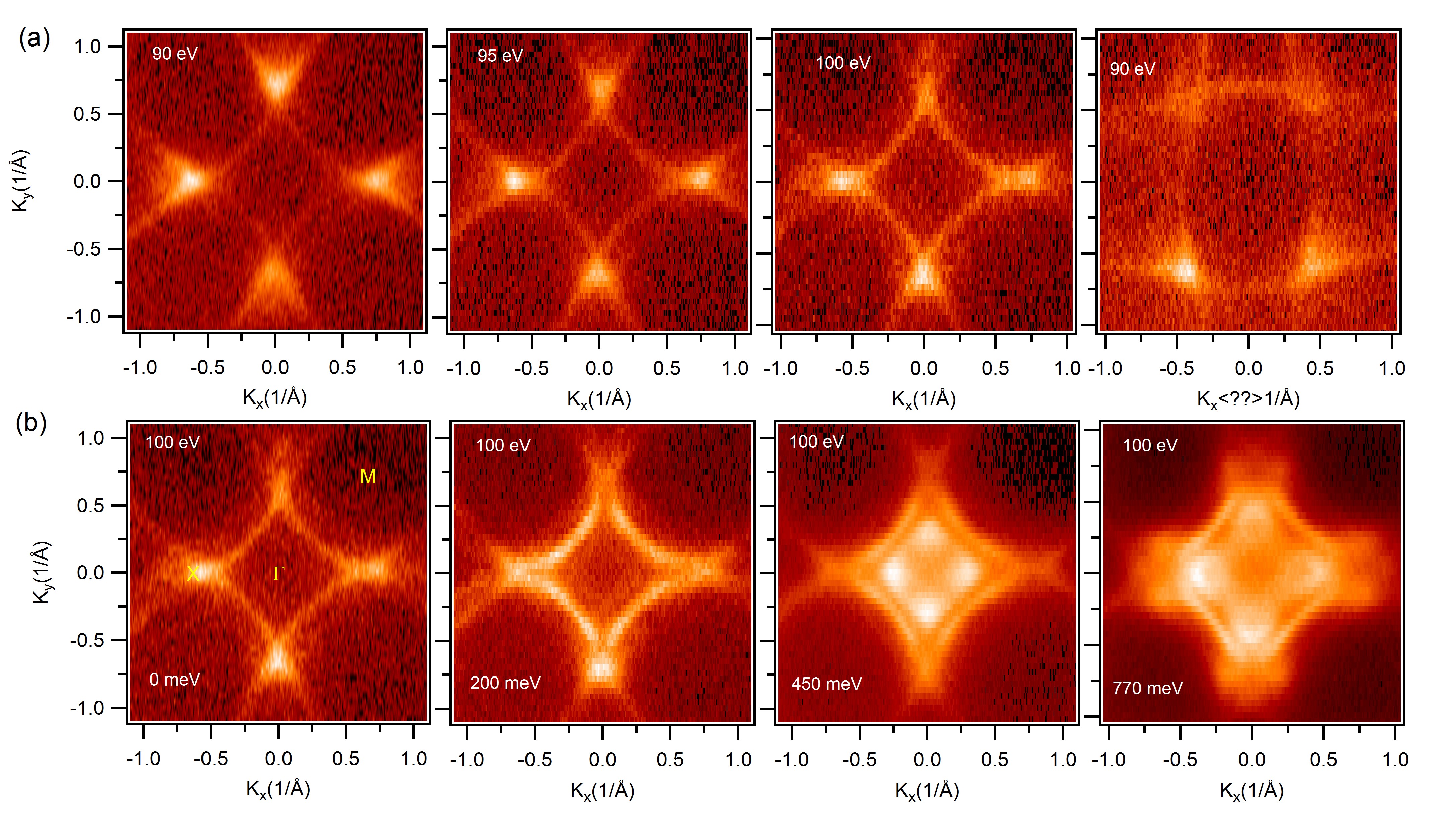}
	\caption{{Fermi surface and constant energy contour plots.} {(a)} Experimentally measured Fermi surface maps at various photon energies and at different high symmetry directions. Photon energies are noted in the plots. (b) Constant energy contour plots at various binding energies. Energies are noted in the plots. High symmetry points are indicated in the leftmost plot. All the measurements were performed at the ALS beamline 4.0.3 at a temperature of 21 K. }
\end{figure*}

Single crystals of GdSbTe were grown by chemical vapor transport method using iodine as a transporting agent \cite{growth, Material}. The chemical composition of the crystals was checked by energy-dispersive X-ray analysis using a FEI scanning electron microscope equipped with an EDAX Genesis XM4 spectrometer. Single crystal X-ray diffraction (XRD) experiment was performed on a Kuma-Diffraction KM4 four-circle diffractometer equipped with a CCD camera using Mo K$\alpha$ radiation. The results indicated the tetragonal space group \textit{P4/nmm} (No. 129) with the lattice parameters a = 4.3104(5) Å and c = 9.1233(15) Å. The XRD scans revealed that the platelet-shaped crystals have their large surface perpendicular to the crystallographic c-axis. Magnetic measurements were performed in the temperature range from 1.72 K to 400 K and in external fields up to 5 T using a Quantum Design MPMS SQUID magnetometer. For this experiment a set of several single-crystalline platelets was oriented perpendicular to the magnetic field direction (B $\|$ c-axis). Synchrotron based ARPES measurement were performed at Advanced Light Source (ALS) beamline 4.0.3 with a Scienta R8000 hemispherical electron analyzer setup. The angular resolution was set to be better than 0.2$^\circ$. And the energy resolution was set to be better than 20 meV. The electronic structure calculations were carried out using the Vienna \textit{Ab-initio} Simulation Package (VASP) \cite{Kress_1, VASP}, and the generalized gradient approximation (GGA) used as the DFT exchange-correlation functional  \cite{Blochl_1}. Projector augmented-wave pseudopotentials \cite{GGA_1} were used with an energy cutoff of 500 eV for the plane-wave basis, which was sufficient to converge the total energy for a given \textit{k}-point sampling. To simulate the surface effects, we used 1 $\times$ 5 $\times$ 1 supercell for the (001) surface, with a vacuum thickness larger than 19  \AA.

We start our discussion by presenting the crystal structure of the GdSbTe. Similar to other ZrSiS type materials it also crystallizes into a non-centrosymmetric tetragonal crystal structure with space group \textit{P4/nmm}. Fig. 1(a) shows the crystal  structure of GdSbTe, where the Gd-Te bilayers are sandwiched between the layers of Sb atoms forming a square net. The sample quality is further examined by performing spectroscopic core-level measurement (Fig. 1(b)). Sharp peaks of Te 4\textit{d} ($\sim$ 40 eV), Sb 4\textit{d} ($\sim$ 33 eV) and Gd 4\textit{f} ($\sim$ 8.5 eV) are observed which confirmed the excellent sample quality used for our measurements. The black dashed line represents the Fermi level. Fig. 1(c) displays the magnetic properties of GdSbTe. Above 15 K, the magnetic susceptibility, $\chi$(T), obeys a Curie-Weiss law with the effective magnetic moment $\mu_{eff}$ = 7.71 $\mu$$_B$ and the paramagnetic Curie temperature  $\theta$= -19 K. The experimental value of $\mu_{eff}$ is close to that expected for a trivalent Gd ion (7.94 $\mu$$_B$). The large negative value of $\theta$ signals strong antiferromagnetic exchange interactions. 
As shown in the upper inset of Fig. 1(c), the compound orders antiferromagnetically at T$_N$ = 13 K, and undergoes another magnetic phase transition near 8 K. At the lowest temperature attained in the present work, i.e. T = 1.72 K, the compound bears an antiferromagnetic state, as signaled by the character of the magnetization isotherm displayed in the lower inset to Fig. 1c. The magnetic field variation of the magnetization, $\sigma$(H), exhibits some tiny inflection near 1.5 T, which can be attributed to a metamagnetic-like phase transition. In stronger fields, $\sigma$(H) of GdSbTe shows a feebly upward behavior, which indicates that an expected field-induced ferromagnetic arrangement of the gadolinium magnetic moments can be achieved in this compound in magnetic fields much stronger than the maximum field of 5 T available in the present study. We performed \textit{ab-initio} bulk band calculations on this system along the various high symmetry directions (Fig. 1(d)). Dirac like feature with a small gap is observed at the X point at around 0.2 eV below the Fermi energy. 

\begin{figure*}
	\centering
	\includegraphics[width=18cm]{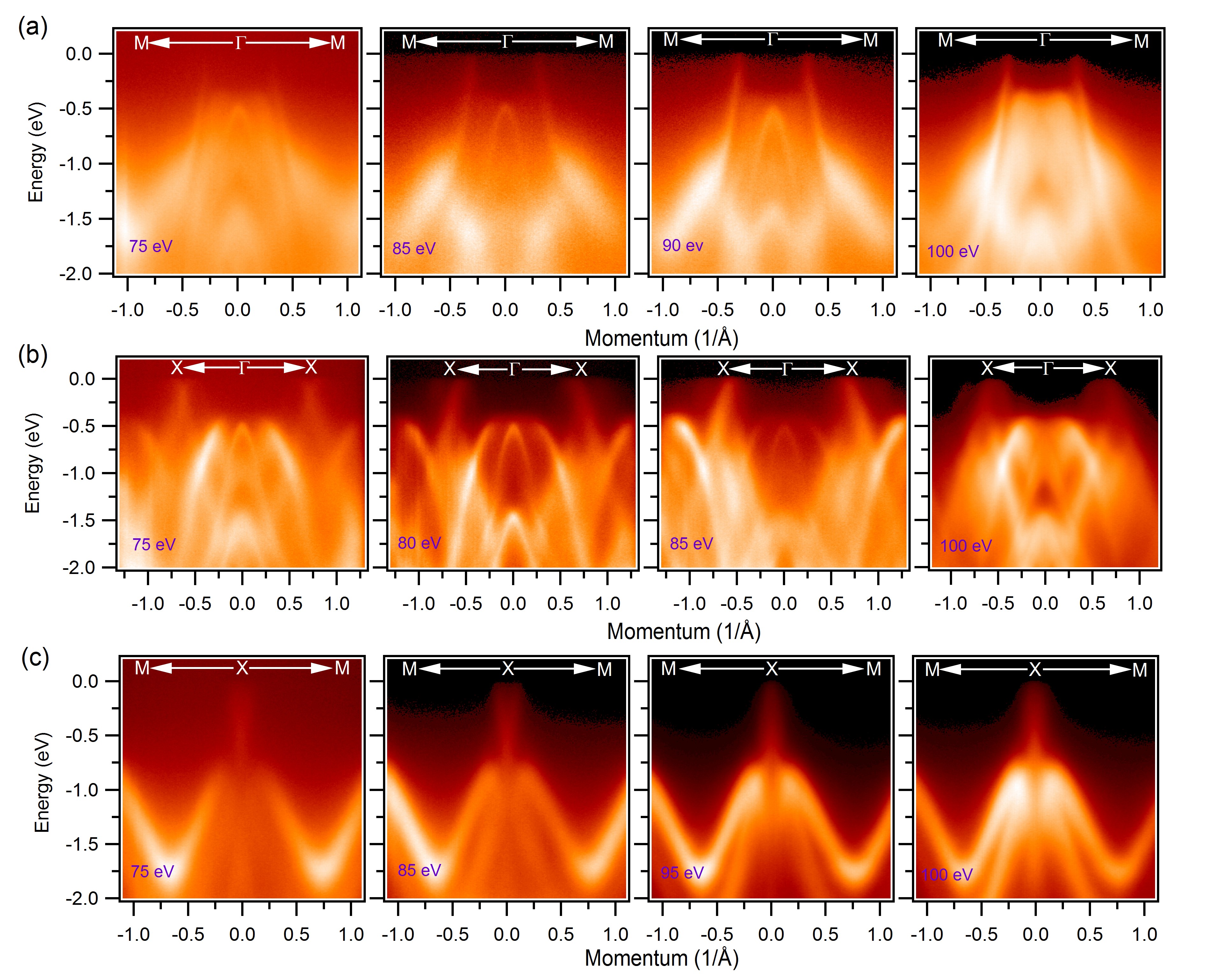}
	\caption{{Dispersion map along the high symmetry directions.} {(a)} ARPES measured dispersion maps along the M-$\Gamma$-M direction at various photon energies. Nodal-line is observed to be in the vicinity of the chemical potential. (b) Dispersion maps along the high symmetry X-$\Gamma$-X direction. (c) Band dispersion along the M-X-M direction. Dirac-like state is observed. Measured photon energies are noted in the plots. All the measurements were performed at the ALS beamline 4.0.3 at a temperature of 21 K. }
\end{figure*}

Figure 2(a) shows Fermi surface maps of GdSbTe in its paramagnetic states at various incident photon energies. High symmetry points and energies are noted in the plots. The overall Fermi surface is similar to previously reported data on ZrSiS-type materials \cite{Neupane_5, ZrSiX}. The topological nodal-line states along the M-$\Gamma$-M directions form a diamond shaped Fermi pockets around the center of the BZ. Furthermore, we observe a small circular and a large elliptical shaped Fermi pockets around the X and M points of the BZ, respectively. Fig. 2(b) shows the constant energy contour plots at various binding energies. Moving towards the higher binding energy we observe that the diamond shape gradually evolves into two diamonds. At a higher binding energy a small circular pocket like feature evolves along the $\Gamma$-X direction. The small pocket is clearly visible at constant energy contour plot with binding energy of 770 meV which is in the vicinity of the Dirac point along this direction.

To reveal the nature of the states along the various high symmetry directions, we present dispersion maps in Figure 3. Fig. 3(a) shows the measured dispersion maps along the M-$\Gamma$-M direction at various photon energies, where Dirac line node phase has been observed. Fig. 3(b) represents the dispersion maps along the X-$\Gamma$-X direction. To fully understand the nature of the state along the X point, we present dispersion maps along the M-X-M high symmetry direction in Fig. 3(c). Photon energies are noted in the plots. The Dirac like state is observed at the X point. In the vicinity of the Dirac point two other bulk bands are observed. 

  \begin{figure*}
  	\centering
  	\includegraphics[width=18cm]{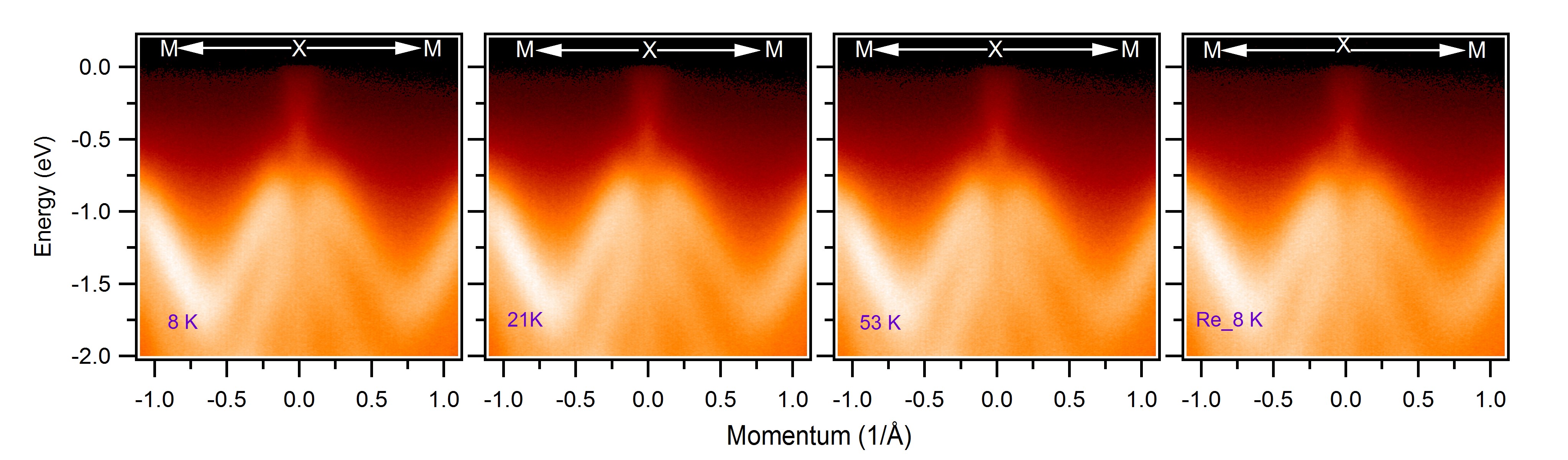}
  	\caption{{Temperature dependent measurement of dispersion maps  along the M-X-M direction.}  Measured temperature are noted on the plots. Re$\_$8K indicates the dispersion map after thermal recycle (8K$\rightarrow$21K$\rightarrow$53K$\rightarrow$8K). All the measurements were performed at the ALS beamline 4.0.3. }
  \end{figure*}


Since the antiferromagnetic (AFM) transition temperature is 13 K in this system, it can provide a platform to study the effect of magnetism and topology. We present dispersion maps along the M-X-M direction both below and above the magnetic transition temperature of this material in Fig. 4. Temperature values are indicated in the plots. Interestingly, we observe the Dirac like state in both paramagnetic and antiferromagnetic phases. Our thermally recycled dispersion map (8 K$\rightarrow$ 21K$ \rightarrow$ 53K $ \rightarrow$ 8K) indicates the robust nature of the Dirac state in GdSbTe. Although the time reversal symmetry (T) is broken in magnetic phase, the product of screw C$_{2x}$ and T might be responsible for protecting the Dirac state in the AFM phase (see Supplementary Information).

In conclusion, we performed detailed studies of GdSbTe material using ARPES, transport measurements and first-principles calculations. Our studies reveal the topological nodal state in this material and confirm that the overall electronic structure is similar with ZrSiS-type materials. Most importantly, we observe a Dirac-like state in both below and above the magnetic transition temperature which is around 13 K. With a relatively higher magnetic transition temperature, GdSbTe can be a potential material to study the interactions between the magnetism and topology.



\bigskip
\bigskip

M.N. is supported by the start-up fund from University of Central Florida.
T.D. is supported by NSF IR/D program. 
D.K. was supported by the National Science Centre (Poland) under research grant 2015/18/A/ST3/00057.
 We thank Jonathan Denlinger for beamline assistance at the LBNL.

\*Correspondence and requests for materials should be addressed to M.N. (Email: Madhab.Neupane@ucf.edu).

\end{document}